\documentclass[11pt]{JHEP3}
\usepackage{amsmath,amssymb,latexsym,cite,graphics,setspace, bbm}

\usepackage[matrix,arrow]{xy}

\def\C{{\mathbb C}}

\def\Z{{\mathbb Z}}

\def\ad{\operatorname{ad}}

\def\tr{\operatorname{tr}}

\def\SO{\operatorname{SO}}

\def\SU{\operatorname{SU}}
\def\GU{\operatorname{U{}}}

\def\GE{\operatorname{E}}

\def\p{\partial}
\def\pb{\bar{\partial}}

\def\ff#1#2{{\textstyle\frac{#1}{#2}}}

\def\cA{{\cal A}}

\def\cD{{\cal D}}

\def\cF{{\cal F}}
\def\cG{{\cal G}}

\def\cI{{\cal I}}
\def\cJ{{\cal J}}
\def\cK{{\cal K}}
\def\cL{{\cal L}}

\def\cN{{\cal N}}
\def\cO{{\cal O}}

\def\cQ{{\cal Q}}

\def\ep{{\epsilon}}

\newcommand\betah{\widehat{\beta}}

\newcommand\omegah{\widehat{\omega}}

\newcommand\alphab{\overline{\alpha}}
\newcommand\betab{\overline{\beta}}

\newcommand\vphi{\varphi}

\newcommand\Omegab{\overline{\Omega}}

\newcommand\gh{\widehat{g}}

\newcommand\cb{\overline{c}}

\newcommand\zb{\overline{z}}

\newcommand\Eh{\widehat{E}}

\newcommand\Hh{\widehat{H}}

\newcommand\Rh{\widehat{R}}

\def\iden{{\mathbbm 1}}

\def\HH#1#2#3{H_{#1~#3}^{~#2}}

\def\bLambda{{\boldsymbol{\Lambda}}}
\def\blambda{{\boldsymbol{\lambda}}}

\def\bcA{{\boldsymbol{\cA}}}
\def\bcAh{{\boldsymbol{\widehat{\cA}}}}
\def\bca{{\boldsymbol{a}}}
\def\bcF{{\boldsymbol{\cF}}}

\def\bcFh{{\boldsymbol{\widehat{\cF}}}}
\def\bcDh{{\boldsymbol{\widehat{\cD}}}}
\def\bQ{{\boldsymbol{Q}}}
\def\bP{{\boldsymbol{P}}}
\def\bR{{\boldsymbol{R}}}

\def\cIh{{\widehat{\cI}}}
\def\cKh{{\widehat{\cK}}}

\def\CO#1#2{{[#1,#2]}}
\def\AC#1#2{{\{#1,#2\}}}

\def\Pic{\operatorname{Pic}}

\title{Heterotic Sigma Models with $N=2$ Space-Time Supersymmetry}
\author{Ilarion V.~Melnikov\\
\normalsize Max-Planck-Institut f\"ur Gravitationsphysik (Albert-Einstein-Institut),\\
\normalsize Am M\"uhlenberg 1, D-14476 Golm, Germany\\
Email:  \email{ilarion@aei.mpg.de}
}
\author{Ruben Minasian\\
\normalsize Institut de Physique Th{\'e}orique, CEA/Saclay \\
\normalsize 91191 Gif-sur-Yvette Cedex, France\\
Email: \email{ruben.minasian@cea.fr}
}

\abstract{We study the non-linear sigma model realization of a heterotic vacuum with N=2 space-time supersymmetry.  We examine the requirements of (0,2) + (0,4) world-sheet supersymmetry and show that a geometric vacuum must be described by a principal two-torus bundle over a K3 manifold.}

\preprint{AEI-2010-155 \\ IPHT-T10/160}

\begin{document}

\section{Introduction}
Flux compactifications occupy a substantial portion of the landscape of current string theory research.
Whether the aims are phenomenological or fundamental, the space-time point of view has been most
prominent.  This is understandable, given the difficulties in formulating a world-sheet
approach suitable for general backgrounds of the type II string.

Heterotic flux compactifications have also been receiving a share of attention.  That such backgrounds exist was already clear based on string duality arguments presented some time ago~\cite{Dasgupta:1999ss}.  More recently, the supergravity equations for compactifications preserving $N=1$ super-Poincar{\'e} invariance in four
dimensions, originally derived in~\cite{Strominger:1986uh}, were solved~\cite{Fu:2006vj} in the context of a specific
$\SU(3)$-structure geometry proposed in~\cite{Goldstein:2002pg}.\footnote{While this work is concerned
with $N=2$ spacetime supersymmetry, the existence of $N=1$ solutions has been explored in, for instance, 
\cite{Li:2004hx,Andreas:2010qh}.}

The supergravity approach is powerful and elegant, especially when formulated in the language of G-structures.
For instance, it was systematically applied in~\cite{Gauntlett:2003cy} to classify the necessary local geometric
conditions for the preservation of various numbers of super-charges in both type II and heterotic contexts.
The world-sheet offers a complementary approach, which is at least in principle more general:  a sufficiently
powerful string theorist would simply study the abstract superconformal two-dimensional theory, with possible
geometric interpretations and supergravity limits emerging as simple corollaries of the SCFT results.  A less 
hypothetical being can start with a non-linear sigma model (NLSM) description and attempt to systematically
study conditions for conformal invariance.  A starting point for such explorations must be the proper identification
of various world-sheet (super)symmetries that should be preserved by the corrections.  

In the context of pertrubative heterotic strings, the requisite symmetries were identified some time 
ago~\cite{Banks:1987cy,Banks:1988yz,Ferrara:1989ud}.  For instance, a necessary and sufficient condition for $N=1$
SUSY in $d=4$ is for the (0,1) superconformal invariance of the heterotic string to be enhanced to (0,2),
with states carrying integral charges under the R-symmetry.  When the SCFT is realized as a NLSM, it is known that to one-loop order in $\alpha'$ the conditions for (0,2) invariance are indeed closely related to
those obtained via supergravity analysis~\cite{Hull:1985jv,Strominger:1986uh,Sen:1986mg}.  We will review the
relation below.

This note is mainly concerned with an application of this idea in the context of $N=2$ heterotic backgrounds,
where the world-sheet theory must possess commuting (0,4) and (0,2) SUSY algebras.  Of course a product 
theory with target-space $K3\times T^2$ obviously possesses such a structure.  What is perhaps more surprising 
from the world-sheet point of view is that there is a more general solution as well.

Assuming that such a background is represented by a NLSM with a smooth geometry, and that the requisite
symmetry of the SCFT is already identifiable in the Lagrangian, we will show that the target-space $X$ must be
a $T^2$ bundle $\pi : X \to B$ over a base $B =K3$ equipped with a conformally hyper-K\"ahler metric.  Moreover, the heterotic
bundle data consists of the pull-back of a stable holomorphic bundle $\Eh \to B$ and a choice of commuting Wilson lines for the $T^2$ directions.  The data must satisfy the topological constraints encoded in the heterotic Bianchi identity.  
These target-spaces are special cases of the manifolds studied in~\cite{Goldstein:2002pg,Fu:2006vj,Becker:2006et,Evslin:2008zm,Swann:2008th,Andriot:2009fp} and are consistent with the supergravity classification results of~\cite{Gauntlett:2003cy}. 
The assumption of a smooth geometry is crucial:  additional theories can be
constructed either as orbifolds of smooth geometries~\cite{Becker:2008rc}, or truly non-geometric theories~\cite{McOrist:2010jw}.

The NLSM we construct is in general strongly coupled:
the target-space necessarily has string-scale cycles for any non-trivial choice of topological data
satisfying anomaly cancellation conditions~\cite{Adams:2009tt}.   Strictly speaking, this means that
neither supergravity nor the NLSM offers a controlled approximation.   Nevertheless, we may hope that the extended space-time
and world-sheet supersymmetries may give a sufficiently rigid structure to constrain possible quantum corrections.  To make this hope into a tangible program, the first step would be to develop a superspace formulation for theories with (0,2)+(0,4) world-sheet supersymmetry.  We leave this as an important open problem.

The lay-out of the article is as follows:  in section \ref{s:02rev} we review the connection between N=1 spacetime and
(0,2) world-sheet supersymmetries; in section \ref{s:N2} we consider the requirements of N=2 space-time supersymmetry and solve them in the NLSM context.  We conclude with a discussion of our results in the context of heterotic compactifications, as well as in the general setting of supersymmetric NLSMs.

\acknowledgments  It is a pleasure to thank J.~Arnlind, G.~Bossard, A.~Degeratu, S.~Fredenhagen, and S.~Theisen for useful discussions.  The work of IVM is supported in part by the German-Israeli Project cooperation (DIP H.52) and the German-Israeli Fund (GIF). RM thanks the Alexander von Humboldt foundation for support.  We thank our respective institutions for hospitality while some of this work was completed.

\section{Warm-up with (0,2) supersymmetry} \label{s:02rev}
We will begin by reviewing some well-known material, with the aim of introducing some notation and explaining our basic strategy.  

A perturbative heterotic string compactification to four-dimensional Minkowski space requires a choice of a (0,1) super-conformal theory with central charge $(c,\cb) = (22,9)$ and a GSO projection consistent with modular invariance.  In a large radius limit of the compactification, if such a limit exists exists, the SCFT is well-approximated by a NLSM with (0,1) supersymmetry.  The field-content of such a theory is most conveniently presented in (0,1) superspace.  The details of this construction are well-known and may be found in, for instance,~\cite{West:1990tg}.

Working with a world-sheet metric of signature $(-,+)$, the superspace coordinates are taken to be $x^-,x^+, \theta^+$.  The superspace covariant derivative $\cD$ and supercharge $\cQ$ are given by
\begin{align}
\cD = \p_{\theta^+} + i \theta^+ \p_+,\qquad \cQ = \p_{\theta^+} -i \theta^+\p_+
\end{align}
and satisfy the (0,1) SUSY algebra:
\begin{align}
\cD^2  = i\p_+, \quad \cQ^2 = -i\p_+,\quad \cQ \cD + \cD \cQ =0.
\end{align}
There are two natural types of superfield:  the matter superfields $\Phi$, containing the bosons $\phi^\mu(x)$, $\mu=1,\ldots,n$, which locally describe the maps from the world-sheet to the target-space manifold $X$ of dimension $n$, as well as their super-partners $\psi_+^\mu$; and $2r$ Fermi superfields $\Lambda$, containing left-moving fermions $\lambda_-$.  The component expansions are:
\begin{equation}
\label{eq:01superfields}
\begin{array}{lcl}
~~\Phi = \phi +  \theta^+ \psi_+ & ~~ & ~~\Lambda = \lambda_- + \theta^+ L \\
\cD \Phi = \psi_+ + i \theta^+\p_+\phi & ~~ & \cD\Lambda = L + i \theta^+ \p_+\lambda_- \\
\cQ \Phi = \psi_+ - i \theta^+\p_+\phi & ~~ & \cQ \Lambda = L - i \theta^+ \p_+\lambda_-
\end{array}
\end{equation}
It will be convenient to combine the left-moving multiplets into a single vector $\bLambda$.

\subsection{A (0,1) heterotic NLSM}
The classically scale-invariant (0,1) supersymmetric action for this field-content is given by
\begin{equation}
S = \frac{1}{4\pi\alpha'} \int d^2 x ~d\theta^+ \left[- i E_{\mu\nu}(\Phi) \cD \Phi^\mu \p_- \Phi^\nu - \bLambda^T D\bLambda \right],
\end{equation}
where
\begin{align}
E_{\mu\nu}(\Phi)   &= g_{\mu\nu}(\Phi) + B_{\mu\nu}(\Phi), \nonumber\\
D \bLambda &= \cD \bLambda + \cD\Phi^\mu \bcA_{\mu }(\Phi) \bLambda.
\end{align}
It will be useful to have the action in components.  We write 
\begin{align}
4\pi \alpha' S = \int d^2 x \left[ \cL_\phi + \cL_{\blambda} \right], 
\end{align}
and carrying out the component expansion find
\begin{align}
\cL_\phi & =  (g_{\mu\nu} + B_{\mu\nu}) \p_+\phi^\mu \p_-\phi^\nu +ig_{\mu\nu} \psi_+^\mu ( \p_-\psi_+^\nu + \p_-\phi^\lambda(\Gamma^\nu_{\lambda\rho} + \ff{1}{2} H^{\nu}_{~\lambda\rho} ) \psi_+^\rho ),\nonumber\\
\cL_{\blambda} & = i \blambda^T (\p_+\blambda + \p_+\phi^\mu \bcA_\mu \blambda)+\ff{1}{2}\psi^\mu\psi^\nu\blambda^T \bcF_{\mu\nu} \blambda,
\end{align}
with $\Gamma$ being the usual Christoffel connection and
\begin{align}
H_{\nu\rho\lambda} &= B_{\nu\rho,\lambda} + B_{\lambda\nu,\rho}+B_{\rho\lambda,\nu}, \nonumber\\
\bcF_{\mu\nu} &= \bcA_{\nu,\mu} - \bcA_{\mu,\nu} + \bcA_\mu\bcA_\nu-\bcA_\nu\bcA_\mu.
\end{align}
Note that we work with anti-Hermititan generators for the gauge fields.
The geometric interpretation is now clear:  the $\phi^\mu$ describe the map from the world-sheet to the target-space $X$, equipped with a metric $g$ and a B-field $B$; the right-moving fermions are sections of $T_X$, coupled to the Christoffel connection twisted by $H = dB$; and the left-moving fermions are sections of a vector bundle $E \to X$ equipped with a connection $\bcA$ with curvature $\bcF$.  By construction, the theory has a (0,1) SUSY algebra with supercharge $\bQ_1$ and right-moving momentum  $\bP \equiv -i \p_+$.  The action on the component fields is simply\footnote{We use the condensed notation $\bQ_1\cdot \phi \equiv \CO{\bQ_1}{\phi}$, $\bQ_1\cdot \psi \equiv \AC{\bQ_1}{\psi}$, etc. }
\begin{align}
\bQ_1 \cdot \phi^\mu = -i\psi^\mu, \qquad \bQ_1 \cdot \psi^\mu =  \p_+\phi^\mu, \qquad \bQ_1 \cdot \blambda =  i\psi^\mu \bcA_\mu \blambda.
\end{align}
The algebra closes to $\bQ_1^2 =\bP$  when we impose the $\blambda$ equations of motion in $\bQ_1^2\cdot \blambda$.

This classical result receives important quantum corrections.  The most basic of these is the anomaly due to the presence of chiral fermions~\cite{Hull:1985jv,Sen:1986mg,Hull:1986xn}:  the path integral measure is not well-defined unless the Pontryagin classes of $T_X$ and $E$ are equal.  When  this topological condition is obeyed, the anomalous transformation of the fermion measure may be cancelled by a gauge transformation of the B-field.  This is, of course, the world-sheet version of the Green-Schwarz mechanism.  For what follows, an important feature of this cancellation is that in order to maintain supersymmetry at the one-loop level, the $H$-field appearing above must be replaced by the gauge-invariant three-form\footnote{
There is an ambiguity in the choice of local counter-terms in defining the one-loop effective action; this ambiguity translates into choices for the connections appearing in $H$~\cite{Sen:1986mg,Hull:1986xn}.  Compatibility with space-time supersymmetry selects out a preferred connection, which leads to important simplifications in the supergravity analysis.  See~\cite{Becker:2009df, Becker:2009zx} for recent discussion and applications. }
\begin{align}
H = d B + \frac{\alpha'}{4} \left( \omega_{3}(\bcA) -\omega_3(\Gamma_{\text{spin}}) \right),
\end{align}
where $\omega_3(A)$ denotes the Chern-Simons form for the connection $A$, and $\Gamma_{\text{spin}}$ is a spin connection for the metric.   The gauge-invariant $H$ satisfies the familiar Bianchi identity
\begin{equation}
dH = \frac{\alpha'}{4} \left(\tr \bcF\wedge \bcF- \tr R\wedge R\right),
\end{equation}
where $R$ is the curvature of the spin-connection.  The resulting NLSM is difficult to study, and not much is known about the general conditions for which it defines a non-trivial SCFT. 

\subsection{Space-time and world-sheet supersymmetries}
A heterotic NLSM is under much better control when it describes a background with $N=1$ space-time supersymmetry.  The reason for this is that a heterotic SCFT describes a (string) perturbative vacuum with $N=1$ super-Poincar\'e invariance in $d=4$ if and only if the SCFT posseses (0,2) world-sheet superconformal symmetry, with the R-charges of all states obeying certain integrality conditions~\cite{Banks:1987cy} which ensure the existence of a well-defined spectral flow operator.  

While this result holds for an arbitrary SCFT, we will apply it in the case that the CFT is realized as a heterotic NLSM.  The simplest way that the CFT can acquire a (0,2) supersymmetry is if
the NLSM Lagrangian already realizes this symmetry.  We will
now review how the requirement of (0,2) supersymmetry
restricts the NLSM.

We seek a theory that realizes the (0,2) algebra given in terms of two supercharges $\bQ_1$, $\bQ_2$, the R-symmetry charge $\bR$, and the right-moving translation generator $\bP$.  The non-trivial commutation relations are
\begin{equation}
\CO{\bR}{\bQ_A} = i\ep_{AB} \bQ_B, \qquad \AC{\bQ_A}{\bQ_B} = 2\delta_{AB} \bP.
\end{equation}
The (0,1) NLSM described above already provides us with a candidate $\bQ_1$ and $\bP$.  The R-symmetry must be realized as a chiral action on the fermions $\psi$: 
\begin{equation}
\bR \cdot \psi^\mu = -i\cJ^\mu_\nu(\phi) \psi^\nu.
\end{equation}
Clearly $\bR$ satisfies $\CO{\bR}{\bP} = 0$.
A short calculation~\cite{Hull:1985jv,Sen:1986mg} shows that R-invariance of
the matter action requires the tensor $\cJ$ to be compatible with the metric and covariantly constant with respect to the twisted connection $\nabla^-$:
\begin{align}
\label{eq:Rsymcond1}
0& = \cJ^\nu_\mu g_{\nu\lambda} + \cJ^\nu_\lambda g_{\nu\mu}, \nonumber\\
0 & = \nabla^-_\nu \cJ^\mu_\lambda = \cJ^{\mu}_{\lambda,\nu} + (\Gamma^{\mu}_{\nu \rho} -\ff{1}{2} H_{\nu~\rho}^{~\mu}) \cJ^{\rho}_{\lambda} 
- (\Gamma^{\rho}_{\nu\lambda} -\ff{1}{2} H_{\nu~\lambda}^{~\rho}) \cJ^{\mu}_{\rho}.
\end{align}
The Fermi action $\cL_{\blambda}$ will be invariant under this R-symmetry if
\begin{equation}
\label{eq:Rsymcond2}
\cJ^\nu_{\mu} \bcF_{\nu\lambda} +\bcF_{\mu\nu}\cJ^\nu_\lambda = 0.
\end{equation}
When the target-space admits such a choice of background fields, we can use the commutation relations to obtain the second supersymmetry:
\begin{align}
\bQ_2\cdot \phi^\mu &= i\CO{\bQ_1}{\bR} \cdot\phi^\mu = i\cJ^\mu_\nu \psi^\nu \nonumber\\
\bQ_2\cdot \psi^\mu &= i\CO{\bQ_1}{\bR} \cdot \psi^\mu = \cJ^\mu_\nu \p_+\phi^\nu + i\cJ^\mu_{\nu,\rho}\psi^\nu\psi^\rho, \nonumber\\
\bQ_2\cdot \blambda &= i\CO{\bQ_1}{\bR}\cdot\blambda = -i\psi^\nu \cJ^\mu_\nu \bcA_\mu \blambda.
\end{align}
We now want to determine whether the defined generators close to the (0,2) algebra.
In general, the algebra need only close up to the equations of motion; indeed, we already observed this to be the case for the (0,1) algebra, where the $\blambda$ equations needed to be used.  In the case of $(0,q)$ supersymmetry the algebra must close on $\phi$ and $\psi$ fields without equations of motion.  This is an important simplification.

Repeated use of the Jacobi identity shows that the generators will satisfy the $(0,2)$ algebra if and only if $\bQ_1 = i\CO{\bR}{\bQ_2}$.  Evaluating this on the fields, we find two conditions:
\begin{align}
\bQ_1\cdot \phi^\mu &= i\CO{\bR}{\bQ_2}\cdot \phi^\mu \implies \cJ^2 = - \iden ,\nonumber\\
\bQ_1\cdot \psi^\mu &= i\CO{\bR}{\bQ_2}\cdot \psi^\mu \implies \cN^\mu_{\lambda\rho} = 0,
\end{align}
where
\begin{equation}
\cN^\mu_{\lambda\rho} = \cJ^\mu_{\nu,[\rho} \cJ^\nu_{\lambda]} - 2\cJ^\nu_{[\lambda} \cJ^\mu_{\rho],\nu}   -\cJ^\mu_\nu \cJ^\nu_{[\lambda,\rho]}.
\end{equation}
The first condition means that $\cJ$ is an almost complex structure on $X$; using this in $\cN$ allows us to express it in a more familiar form:
\begin{equation}
\cN^{\mu}_{\lambda\rho} = \cJ^\nu_\lambda (\cJ^\mu_{\rho,\nu}-\cJ^\mu_{\nu,\rho})  - \cJ^\nu_\rho (\cJ^\mu_{\lambda,\nu}-\cJ^\mu_{\nu,\lambda}).
\end{equation}
This is the Nijenhuis tensor for $\cJ$, and its vanishing implies that $\cJ$ defines a complex structure on $X$.

Evidently, (0,2) SUSY requires $X$ to be a complex manifold equipped with a Hermitian form $\omega_{\mu\nu} = \cJ^{\lambda}_\mu g_{\lambda \nu}$.   Moreover (\ref{eq:Rsymcond2}) implies that $E$ must be a holomorphic bundle equipped with a Hermitian connection with a $(1,1)$ field-strength $\bcF$.  The vanishing of $\nabla^-\cJ$ implies just one additional condition on the background~\cite{Strominger:1986uh}:
\begin{equation}
-H_{\mu\nu\rho} =
 \cJ^\lambda_\mu \nabla_\lambda \omega_{\nu\rho}
        +\cJ^\lambda_\rho \nabla_\lambda \omega_{\mu \nu}
        +\cJ^\lambda_\nu \nabla_\lambda \omega_{\rho\mu} .
\end{equation}
This may be written in terms of the Dolbeault operators $\p,\pb$ as $H= i (\p-\pb) \omega$.\footnote{Note that our $B$ and $H$ differ by a sign from conventions common in the supergravity literature~\cite{Gauntlett:2003cy,Becker:2006et}.}

These classical considerations receive important quantum corrections at one loop in $\alpha'$.  First, as discussed above, the $H$ appearing in the one-loop effective action is naturally the gauge-invariant field-strength.  Combining this with form of $H$ in terms of $\omega$, we find the condition
\begin{equation}
\label{eq:Bianchiomega}
i\p\pb \omega = \frac{\alpha'}{8} \left(\tr R\wedge R- \tr \bcF\wedge \bcF\right).
\end{equation}
In addition, the chiral R-symmetry suffers from an anomaly proportional to $c_1(T_X)$~\cite{Hull:1985jv}.  Thus, to maintain (0,2) SUSY $X$ must have $c_1(T_X) = 0$.    In addition, the vanishing of a global anomaly requires $c_1(E) \in H^2(X, 2 \Z)$~\cite{Distler:1986wm}.

In order to construct a space-time supersymmetry generator, the theory must possess a right-moving spectral flow operator.  In a
non-linear sigma model, the square of the spectral flow operator
is given by $\Sigma^2 = \Omega_{\lambda\mu\nu} \psi^\lambda\psi^\mu\psi^\nu$~\cite{Sen:1986mg}.   $\Sigma^2$ must have R-charge $3$, which implies that $\Omega$ is a $(3,0)$ form.  Finally, on-shell $\Sigma^2$ must be a free right-moving field, i.e. $\p_- \Sigma^2 = 0$.\footnote{Actually, another necessary condition is that $\Sigma^2$ must be a chiral operator, i.e. it must be annihilated by $Q_1+iQ_2$.  It is not hard to show that this condition is satisfied when $\Omega$ is a (3,0)-form.}  Using the equations of motion for $\psi$ we find 
\begin{align}
\p_-\Sigma^2 = \ff{3i}{2} \Omega_{\alpha\mu\nu} g^{\alpha\gamma} \blambda^T \bcF_{\gamma\lambda}\blambda 
\psi^\lambda \psi^\mu\psi^\nu 
+ \nabla^-_\beta \Omega_{\lambda\mu\nu} \p_-\phi^\beta \psi^\lambda\psi^\mu\psi^\nu.
\end{align}
The two terms must vanish separately, leading to two constraints on the geometry.  The first term requires $\Omega_{\alpha[\lambda\mu} \bcF^\alpha_{\nu]} = 0$.  Writing this in complex coordinates, it is easy to see that $\bcF$ must not only be a $(1,1)$ form, but also satisfy the zero-slope Hermitian Yang-Mills (HYM) equations:  $\omega\wedge \omega \wedge \bcF = 0$.  The vanishing of the second term requires $\nabla^- \Omega = 0$,
so that $\nabla^-$ has $\SU(3)$ holonomy, and in particular $c_1(T_X) = 0$.  It is easily seen that $||\Omega||^2$ is constant on $X$, and furthermore there exists a real closed 1-form $\beta$ such that
\begin{align}
\label{eq:beta6}
d \Omega = \beta \wedge \Omega, \qquad d\Omegab = \beta \wedge \Omegab, \qquad 
d (\omega\wedge \omega) = \beta \wedge \omega\wedge \omega.
\end{align}

So far, we have seen that the world-sheet conditions for $N=1$ space-time SUSY require $X$ to be a manifold
with $\SU(3)$ structure.  From the supergravity point of view~\cite{Strominger:1986uh}, we know that one condition is still missing:  the vanishing of the dilatino variation.  This is equivalent to the closure of 
$\omega$ being conformally balanced by the dilaton field $\vphi$~\cite{Li:2004hx,Becker:2006et}:
\begin{align}
\label{eq:conban}
d (e^{-2\vphi} \omega \wedge \omega ) = 0.
\end{align}  
Comparing to~(\ref{eq:beta6}), we see that $\beta = 2 d\vphi$.  Since $\beta$ is closed, this does not impose any condition on the local geometry; however, if we wish the dilaton to be single-valued
on $X$, then we must demand that $\beta$ is exact.  This is an additional topological requirement, since
in this case $e^{-2\vphi} \Omega$ is a closed, nowhere vanishing (3,0)-form.  This means that $X$ has
a holomorphically trivial canonical bundle, i.e. $h^{(3,0)} (X) = 1$. 

Perhaps it is useful to recall the distinction between topological and holomorphic triviality of line bundles.
Recall that holomorphic line bundles on $X$ are classified by $\Pic(X) = H^1(X,\cO^\ast)$, a sheaf
cohomology group determined by the exact sequence 
\begin{align}
\xymatrix{0 \ar[r] & H^1(X,\Z)  \ar[r] & H^1(X,\cO) \ar[r] & \Pic(X) \ar[r]^{c_1} & H^2(X,\Z) \ar[r] & H^2(X,\cO) \ar[r] &\cdots .}
\end{align}
Thus, we see that elements of $\Pic^0(X) \equiv H^1(X,\cO) /H^1(X,\Z) \subset \Pic(X) $ correspond to isomorphism classes of holomoprhic line  bundles on $X$ with $c_1 = 0$.  While these are all trivial as $C^{\infty}$ bundles, 
they may not be trivial as holomorphic bundles~\cite{Griffiths:1978pa}.  We recognize $\Pic^0(X)$ as characterizing the holomorphic Wilson lines on $X$.  There are many examples of non-K\"ahler 
complex manifolds with $c_1=0$ but non-trivial canonical bundle; for instance, even-dimensional Lie groups and the Hopf surfaces described below are simple examples.

In principle, we should be able to recover the conformal balance condition directly from the world-sheet.
Presumably it should arise from an examination of the closure of the full (0,2) algebra, as well as the
OPEs of $\Sigma^2$.  This is not an entirely straightforward undertaking since the dilaton is not so easy
to see on a flat world-sheet, but a careful analysis of the superconformal algebra in conformal gauge should
be feasible.

To summarize, (0,2) SUSY at one-loop in $\alpha'$ requires $E \to X$ to be a holomorphic bundle over a complex manifold $X$ with $c_1(T_X) =0$, equipped with a Hermitian form $\omega$ and connection $\bcA$ constrained by the Bianchi identity of (\ref{eq:Bianchiomega});  there will be a candidate operator for a  space-time supercharge provided that $X$ is an $\SU(3)$ structure manifold and the connection $\bcA$ satisfies the zero-slope HYM equations; finally, the background will admit a single-valued dilaton if $\omega$ is conformally balanced.  

Some of these conditions will receive $\alpha'$ and string corrections.  While we expect that the topological conditions will be unaffected by quantum corrections, the equations for the background fields will be corrected.  For instance, experience with (2,2) Calabi-Yau compactifications suggests that even in $\alpha'$ perturbation theory $\nabla^-$ may no longer be an $\SU(3)$ holonomy connection.  Furthermore, for bundles of non-zero degree a non-zero slope can be generated at one loop in string perturbation theory~\cite{Blumenhagen:2005ga}. 

\section{World-sheet supersymmetry in $N=2$ torsional backgrounds}\label{s:N2}
In the preceding section we reviewed the connection between $N=1$ space-time supersymmetry and (0,2) super-conformal invariance on the world-sheet.  Similar results hold for heterotic compactifications with $N=2$ space-time supersymmetry in four dimensions~\cite{Banks:1988yz,Ferrara:1989ud}.  The result is that $N=2$ space-time supersymmetry is preserved if and only if the right-moving Virasoro algebra is enhanced to a product of a $\bar{c}=6$ (0,4) and a free $\bar{c}=3$ (0,2) superconformal algebra, where the latter is
equipped with a pair of commuting bosonic (non-R) currents.  If these are to be realized in the NLSM, they must correspond to two commuting isometries of the metric and $H$.  Thus we can already conclude that $X$ must be a $T^2$ fibration over a four-dimensional base.

In the context of a NLSM compactification, a well-studied example is the heterotic string on $K3\times T^2$.   In this case the NLSM consists of two decoupled theories, and it is easy to identify the (0,4) and (0,2) supersymmetries.  It is not so clear how to construct this large supersymmetry in the more general case of a torsional background on $X$ constructed as a non-trivial $T^2$ fibration over $K3$.  Arguments based on a dual M-theory description~\cite{Dasgupta:1999ss}, as well as a direct supergravity analysis~\cite{Goldstein:2002pg, Fu:2006vj, Becker:2006et} show that such $N=2$-preserving backgrounds exist.  In this section we will find the requisite world-sheet supersymmetry structures in the NLSM.

\subsection{The desired algebra}
We are interested in NLSMs that possess a (0,2)+(0,4) SUSY.  We will denote this algebra $\cA_2\oplus\cA_4$, with
$\cA_2$ having generators $q_A$, $r$, $p$ and non-trivial commutation relations
\begin{align}
\CO{r}{q_A} = i \ep_{AB} q_B, \quad \AC{q_A}{q_B} = 2\delta_{AB} p.
\end{align}
The (0,4)  algebra $\cA_4$ has a richer structure.  In the standard
presentation, e.g. in~\cite{Green:1987mn}, the supercharges are 
taken to be two doublets under the $\SU(2)$ R-symmetry.  
We will find it
convenient to use the representation which naturally arises from 
the N=4 NLSM construction~\cite{AlvarezGaume:1981hm,Howe:1988cj}.  Taking the $\SU(2)$ R-symmetry generators to be $R_a$
and the four supercharges $Q_0$, $Q_a$, the non-vanishing
commutators are
\begin{align}
\CO{R_a}{R_b} & = 2i \ep_{abc} R_c, \quad
\CO{R_a}{Q_0}  = iQ_a, \quad
\CO{R_a}{Q_b} = -i \delta_{ab} Q_0 + i \ep_{abc} Q_c, \nonumber\\
\AC{Q_a}{Q_b} &= 2 \delta_{ab} P, \quad Q_0^2 = P.
\end{align}

It will be convenient for us to make a choice of a diagonal subalgebra
$\cA_2^{+} \subset \cA_2 \oplus \cA_4$.  Such a choice is
of course not unique, but the ambiguity just corresponds to choosing
an $N=1$ subalgebra of the space-time $N=2$ theory.  Without
loss of generality we will take $\cA_2^{+}$ to be generated by
\begin{align}
\bR = r+R_3, \quad \bQ_{ 1} = q_1 + Q_0, \quad \bQ_2 = q_2 + Q_3,
\quad \bP = p+P,
\end{align}
with the action on the NLSM matter fields as reviewed above\footnote{
Note that for any SUSY transformation, the transformation of the left-moving fermions is determined by
the tranformation of the bosons, e.g.
$\bQ_1 \cdot \blambda =  -(\bQ_1 \cdot \phi^\mu) \bcA_\mu \blambda.$}
\begin{align}
\bP &= -i\p_+, & \bR \cdot \psi^\mu & = -i \cJ^\mu_\nu \psi^\nu, &~ \nonumber\\
\bQ_1 \cdot \phi^\mu &= -i\psi^\mu, & 
\bQ_2\cdot \phi^\mu &= i\cJ^\mu_\nu \psi^\nu,   & \\
\bQ_1 \cdot \psi^\mu &= \p_+\phi^\mu,  & 
\bQ_2\cdot \psi^\mu &= \cJ^\mu_\nu \p_+\phi^\nu+i \cJ^{\mu}_{\nu,\rho}\psi^\nu\psi^\rho.
\end{align}
Closure of $\cA_2^+$ requires $\cJ$ to be a complex structure on $X$, and the action is $\cA^{+}_2$-invariant when (\ref{eq:Rsymcond1}, \ref{eq:Rsymcond2}) hold.  

In order to enlarge the symmetry algebra from $\cA^{+}_2$ to  $\cA_2\oplus\cA_4$,  we must find the generators of the $\GU(1)\times\SU(2)$ R-symmetry $r$, $R_a$.  Assuming the R-symmetries continue to leave the $\blambda$ and $\phi$ invariant,  their action is specified by the four tensors $\cI$, $\cK_a$:
\begin{align}
\label{eq:Rgen}
r \cdot \psi^\mu = -i \cI^{\mu}_\nu \psi^\nu, \quad 
R_a \cdot \psi^\mu = -i\cK^{\mu}_{a\nu} \psi^\nu.
\end{align}
These R-symmetry generators obviously commute with $\bP = -i\p_+$. From $\bR = r + R_3$ we see that $\cJ = \cI + \cK_3$.  
Having found such tensors, we can unambiguously define the generators of $\cA_2\oplus \cA_4$ in terms of those of $\cA_2^+$ and the R-symmetry:
\begin{align}
\label{eq:defs}
q_2 &= -i \CO{r}{\bQ_1}, &
q_1 &= i  \CO{r}{\bQ_2},  & p &= q_2^2,\nonumber\\
Q_a &= -i \CO{R_a}{\bQ_1}, &
Q_0 &= \bQ_1 - q_1, & P &= \bP - p.
\end{align}
More explicitly, we have rather simple expressions for $Q_a$ and $q_2$:
\begin{align}
\label{eq:Qaq2}
Q_a \cdot \phi^\mu &= i\cK^\mu_{a\nu} \psi^\nu, &
Q_a \cdot \psi^\mu & = \cK^\mu_{a\nu} \p_+\phi^\nu +i \cK^\mu_{a\nu,\rho}\psi^\nu\psi^\rho,
\nonumber\\
q_2 \cdot \phi^\mu & = i\cI^\mu_\nu \psi^\nu, &
q_2 \cdot \psi^\mu & = \cI^\mu_{\nu} \p_+\phi^\nu +i \cI^\mu_{\nu,\rho}\psi^\nu\psi^\rho,
\end{align}
while $q_1$ is a bit more complicated:
\begin{align}
\label{eq:q1}
q_1 \cdot \phi^\mu & = i \cJ^\mu_\nu \cI^\nu_\lambda \psi^\lambda,  \qquad
q_1 \cdot \psi^\mu  = -\cI^\mu_\nu \cJ^\nu_\lambda\p_+\phi^\lambda
+ i M^\mu_{\lambda\rho} \psi^\lambda\psi^\rho,\nonumber\\
M^\mu_{\lambda\rho} & = \cJ^\mu_{\nu,[\rho} \cI^\nu_{\lambda]} + \cI^\nu_{[\rho}\cJ^\mu_{\lambda],\nu} 
     -\cJ^\nu_{[\lambda}\cI^\mu_{\rho],\nu}  - \cI^\mu_\nu \cJ^\nu_{[\lambda,\rho]}.  
\end{align}
These transformations will generate symmetries of
the action if $\cI, \cK_a$ are R-symmetries, i.e. (\ref{eq:Rsymcond1}) and
(\ref{eq:Rsymcond2}) are obeyed with $\cJ$ replaced by $\cI$ or $\cK_a$.   

Thus, the transformations are determined by the tensors $\cI$ and $\cK_a$, and our next task is to find the conditions on $\cI$ and $\cK_a$ under which the transformations close to $\cA_2\oplus \cA_4$.  The full list of commutators to be
checked might appear slighlty daunting, but since
some of the relations follow from the others by the Jacobi identity, the computation is tractable.

We begin with a simplifying observation:
there exists another natural (0,2) subalgebra $\cA_2^- \subset \cA_2\oplus\cA_4$
with generators
\begin{align}
\bR' = -r+R_3, \quad \bQ_{ 1}' = \bQ_1, \quad \bQ_2' = -q_2 + Q_3,
\quad \bP' =\bP.
\end{align}
Since $\cA_2^-$ contains the linearly realized (0,1) subalgebra generated by $\bQ_1, \bP$, 
we see that closure of $\cA_2^-$ requires $\cJ_- = - \cI + \cK_3$ to be a second integrable
complex structure on $X$. It is not hard to show that the generators defined in (\ref{eq:Rgen}) 
and (\ref{eq:defs}) close to $\cA_2\oplus\cA_4$ provided that $\cI, \cK_a$ can be chosen so 
that $\cA_2^\pm$ close
(i.e. $\cJ_\pm =\pm \cI +\cK_3$ are complex structures) and the realization satisfies
\begin{align}
\label{eq:A2A4req}
\CO{r}{R_a} & = 0, \qquad \CO{R_a}{R_b} = 2i\ep_{abc} R_c,
 \nonumber\\
\CO{R_a}{q_A} &= 0, \qquad \CO{r}{q_1} = i q_2, \qquad
\CO{R_a}{Q_b}  + \CO{R_a}{Q_b}=0, \quad a\neq b.
\end{align}
When evaluating these requirements on the matter fields, we naturally meet two types of
terms: the first involve various algebraic combinations of the $\cI$ and $\cK_a$ contracted
into either $\psi^\mu$ or $\p_+\phi^\mu$; the second involve the tensors and their derivatives
contracted into a fermion bilinear $\psi^\lambda\psi^\rho$.  Since the two types of terms clearly
do not mix, we may first evaluate the algebraic conditions and then move on to the differential
ones.

\subsection{Algebraic conditions}
The closure of the R-symmetry (first line in (\ref{eq:A2A4req})) leads just to algebraic conditions:
\begin{align}
\label{eq:alg1}
\CO{\cI}{\cK_a} = 0, \qquad \CO{\cK_a}{\cK_b} = 2\ep_{abc} \cK_c.
\end{align} 
The closure of $\cA_2^\pm$ requires $\cJ_\pm^2 = -\iden$, which implies
\begin{align}
\label{eq:alg2}
\cI^2+\cK_3^2 = -\iden, \qquad  \AC{\cI}{\cK_3} = 0.
\end{align}
The remaining algebraic requirements arise from
\begin{align}
\label{eq:alg3}
\CO{R_a}{q_2} = 0 &\implies \cI \cK_a = \cK_a \cI = 0, \nonumber\\
\CO{R_a}{Q_b}+\CO{R_b}{Q_a} = 0, \quad a\neq b & \implies \cK_a \cK_b = \ep_{abc} \cK_c, \quad
a\neq b.
\end{align}
When combined with $\CO{\cK_a}{\cK_b} = 2\ep_{abc} \cK_c$, the latter condition yields
\begin{align}
\label{eq:alg4}
\cK_a \cK_b = \delta_{ab} \cK_3^2 + \ep_{abc} \cK_c.
\end{align}
The remaining relation, $\CO{r}{q_1} = iq_2$, does not lead to additional algebraic constraints.

The algebraic conditions become quite stringent when combined with the metric compatibility conditions (first line of eqn.~(\ref{eq:Rsymcond1})) for $\cI$, $\cJ$ and $\cK_a$ and the torus isometries.  The most general metric on $X$ compatible with the isometries is
\begin{align}
ds^2 = \gh_{ij} dy^i dy^j + \cG_{IJ} (d\theta^I + A^I_i dy^i) (d\theta^J + A^J_j dy^j),
\end{align}
where $i,j= 1,\ldots, 4$, $I,J = 1,2$, and all tensors are independent of the fiber coordinates $\theta^1$ and $\theta^2$. Clearly the metric has the Kaluza-Klein gauge invariance $\theta^I \to \theta^I+f^I(y)$, $A^I \to A^I-d f^I$.  Up to an over-all scaling by a $y$-dependent function, the most general Hermitian form compatible with this gauge invariance is
\begin{equation}
\label{eq:omega}
\omega = \ff{1}{2} \omegah_{ij} dy^i \wedge dy^j +\chi_I \Theta^I
+ \Theta^1\wedge \Theta^2,
\end{equation}
where
\begin{align}
\Theta^I = d\theta^I + A^I_i dy^i
\end{align}
and $\chi_I$ are one-forms on the base.  These one-forms, if non-zero, lead to a mixing between the base and fiber fermions under the R-symmetries.  In what follows, we set $\chi_I = 0$.\footnote{In an earlier version of this paper the $\chi_I\theta^I$ term in $\omega$ was missed.  At this point in the analysis nothing constrains the $\chi_I$ to vanish.  However, as shown in~\cite{Melnikov:2012hf}, these terms must vanish to preserve the (0,2)+(0,4) structure.}

Compatibility of the metric and complex structure determines $\cJ^\nu_{\mu} = \omega_{\mu\rho} g^{\rho\nu}$, and
splitting up the tensor in block form we find
\begin{align}
\cJ
=
\begin{pmatrix}
\cKh^i_{3 j} & ~~ & 0 \\ 
 \cIh^I_M A^M_j - A^I_m \cKh^m_{3 j} & ~~ & \cIh^I_J
\end{pmatrix}.
\end{align}
We have defined $\cKh^i_{3j} = \omegah_{jk} \gh^{ki}$ and $\cIh^I_J = \ep_{JK} \cG^{KI}$.  $\cJ^2 = -\iden$ if and only if $\cKh_3$ and $\cIh$ define almost complex structures in the base and fiber directions, respectively.  

It is easy to see that the algebraic conditions and metric compatibility are satisfied by 
\begin{align}
\label{eq:algsol}
\cI = \begin{pmatrix} ~0 & 0 \\ \cIh A & \cIh \end{pmatrix}, \qquad
\cK_a = \begin{pmatrix} ~~\cKh_a & 0 \\ -A \cKh_a & 0 \end{pmatrix},
\end{align}
when the $\cKh_a$ obey
\begin{align}
\label{eq:Kbase}
\cKh_a \cKh_b = -\iden_4 \delta_{ab} + \ep_{abc} \cKh_c, \quad\text{and}\quad
\cKh^k_{ai} \gh_{kj} + \gh_{ik} \cKh^k_{aj} = 0.
\end{align}
With a little more work it is possible to show that this solution is unique up to diffeomorphisms.\footnote{A straight-forward way to show this is to solve the algebraic requirements at a point in $X$.  The result is that $\cI, \cJ, \cK$ are determined up to an orthogonal transformation.}

Thus, the algebraic conditions require the base manifold to admit an almost hyper-Hermitian structure given 
by the $\cKh_a$ and $\gh$, while $\cIh$ is a metric-compatible complex structure in the fiber directions.  We
parametrize the latter in a standard way in terms of a single complex, possibly base-dependent, parameter
$\tau(y) = \tau_1(y)+i\tau_2(y)$ with $\tau_2 \ge 0$:
\begin{align}
\cIh = \frac{1}{\tau_2} \begin{pmatrix} -\tau_1 & -|\tau|^2 \\ 1& \tau_1 \end{pmatrix}.
\end{align}
The fiber metric $\cG_{IJ}$ then takes the form
\begin{align}
\label{eq:cG}
\cG = \frac{e^{-2\eta(y)}}{\tau_2} 
\begin{pmatrix} 1 & \tau_1 \\ \tau_1 & |\tau|^2 \end{pmatrix}.
\end{align}

\subsection{Differential conditions}
Having found a general solution to the algebraic conditions, we move on to the differential
ones arising from fermion bilinear terms in the transformations.  Our first set of conditions
comes from closure of the $\cA_2^\pm$ subalgebras.  From our discussion of (0,2)
SUSY, it is clear that the differential conditions require $\cJ_\pm$ to be integrable
complex structures.  Splitting the Nijenhuis tensors into the base and fiber components,
we find three non-trivial components, leading to the following requirements:
\begin{align}
\cN^{k}_{ij} (\cJ_\pm)  = 0 &\implies \cN^k_{ij} (\cKh_3) = 0, \nonumber\\
\cN^{K}_{iI} (\cJ_\pm) = 0 &\implies \cKh^m_{3i} \cIh^K_{I,m} \pm \cIh^M_I \cIh^K_{M,i} = 0, \nonumber\\
\cN^{K}_{ij} (\cJ_\pm) = 0 &\implies
F^K_{ij} -\cKh^m_{3i} F^K_{mn} \cKh^n_{3j} \pm \cIh^K_M (\cKh^m_{3i} F^M_{mj} + F^M_{im} \cKh^m_{3j}) = 0.
\end{align}
The first simply means that $\cKh_3$ is an integrable complex structure on the base.  The second condition with
the (-)$+$ sign requires $\tau$ to vary (anti)holomorphically with respect to $\cKh_3$.  The last condition takes
a familiar form when written in complex coordinates $(z^\alpha,\zb^{\alphab})$ on the base:
$F^1_{\alphab\betab} \pm \tau F^2_{\alphab\betab} = 0.$  Consequently, integrability of both $\cJ_+$ and $\cJ_-$ requires $\tau$ to be constant and $F^M$ to be (1,1) forms on the base, i.e.
\begin{align}
\cKh^m_{3i} F^M_{mj} + F^M_{im} \cKh^m_{3j} = 0.
\end{align}
Using these constraints, we obtain a simplification of the $q_1$ transformation: the only non-vanishing
components of the seemingly complicated $M^\mu_{\lambda\rho}$ in (\ref{eq:q1}) are 
\begin{align}
M^K_{ij} = -\ff{1}{2} F^K_{ij}.
\end{align}
The remaining differential conditions for algebra closure arise from 
\begin{align}
\label{eq:KhF}
\CO{R_a}{q_2} \cdot \psi^\mu = 0& \implies \cKh^m_{ai} F^M_{mj} + F^M_{im} \cKh^m_{aj} = 0, \nonumber\\
\CO{R_a}{Q_b} + \CO{R_b}{Q_a} = 0, \quad a\neq b &\implies \cKh_a~~\text{are integrable}.
\end{align}
The latter condition deserves a word of explanation.  When discussing almost hyper-complex structures
 it is convenient to introduce the Nijenhuis
concomitant tensors\footnote{These objects and
their relation to (p,q) world-sheet supersymmetry are
nicely reviewed in~\cite{Howe:1988cj}.} for the $\cKh_a$ defined by
\begin{align}
\cN^{k}_{ij}(\cKh_a,\cKh_b) & = 
\left\{ \cKh^m_{a i} (\cKh^k_{bj,m}-\cKh^k_{bm,j}) + (a\leftrightarrow b) \right\} 
- (i\leftrightarrow j).
\end{align}
For $a=b$ this reduces to (twice) the usual Nijenhuis tensor for an almost complex 
structure $\cKh_a$.  It can be shown that if any two of these concomitants vanish, then
all are zero and the manifold is hypercomplex.  The differential condition we obtain by
direct computation of the commutator is 
$\cN(\cKh_a,\cKh_b) = 0$ for $a\neq b$; however this is equivalent to the integrability of $\cKh_a$.  

Finally, we must ensure that the R-symmetries generated by $\cI$ and $\cK_a$ really are 
symmetries of the action.  This, combined with the (0,1) SUSY, is sufficient to
ensure that we have the full $\cA_2\oplus\cA_4$ symmetry.  We have already discussed
the metric compatibility requirements among our algebraic conditions.  To keep the matter
action invariant, we must also satisfy the differential condition in (\ref{eq:Rsymcond1}) for
$\cI$ and $\cK_a$:
\begin{align}
\label{eq:NIJ}
\nabla_\nu \cI^\mu_\lambda = \ff{1}{2} (\HH{\nu}{\mu}{\rho} \cI^\rho_\lambda - \HH{\nu}{\rho}{\lambda} \cI^\mu_\rho), \qquad
\nabla_\nu \cK^\mu_{a\lambda} = \ff{1}{2} (\HH{\nu}{\mu}{\rho} \cK^\rho_{a\lambda} - \HH{\nu}{\rho}{\lambda} \cK^\mu_{a\rho}).
\end{align}
Expanding (\ref{eq:NIJ}) in base and fiber components, we find several
conditions.  First, $\cG$ is constant, so that without loss of generality we may set $\eta(y) = \text{constant}$ in (\ref{eq:cG}).  Second, the components
of $H$ with legs in the fiber directions are given by
\begin{align}
H_{IJk} = 0, \qquad H_{Ijk} = \cG_{IJ} F^J_{jk}.
\end{align}
Finally, we must have
\begin{align}
H_{ijk} = \Hh_{ijk} + \cG_{MN} \left[A^M_{i} F^N_{jk} + A^M_{k} F^N_{ij} + A^M_{j} F^N_{ki}\right],
\end{align}
where the $3$-form on the base $\Hh$ is given by
\begin{align}
\label{eq:HK}
-\Hh_{ijk} = 
\cKh^m_{ai} \nabla_m \cKh_{ajk} +
\cKh^m_{ak} \nabla_m \cKh_{aij} +
\cKh^m_{aj} \nabla_m \cKh_{aki} \quad \text{(no sum on $a$)}.
\end{align}
We will see shortly that (\ref{eq:HK}) determines $\Hh$ and does not lead to any further conditions on the geometry.

\subsection{Hyper-Hermitian surfaces and $N=2$ backgrounds}
The conditions we have uncovered so far imply that the target-space $X$ is a $T^2$ bundle
over a hyper-complex surface $B$.  In fact, since the $\cKh_a$ are also required to be compatible with the base metric $\gh$, $B$ is actually hyper-Hermitian.
Such complex surfaces are rather well-understood,\footnote{Hyper-Hermitian manifolds are reviewed in~\cite{Joyce:2007rh}, as well as in a nice Wikipedia article.} and we will here review the pertinent results.  The first point is that the Hermitian
forms $\omegah_{aij} = \cKh^k_{ai} \gh_{kj}$ satisfy
\begin{align}
\label{eq:beta}
d \omegah_a = \betah \wedge \omegah_a,
\end{align}
where $\betah$ is a closed one-form determined solely by the
metric.  A short computation shows that (\ref{eq:HK}) is equivalent to
$\Hh = -\ast_4 \betah$, showing that $\nabla^- \cI = \nabla^-\cK_a = 0$ determine the torsion and do not place any additional constraints on the geometry.

Compact hyper-Hermitian surfaces are classified~\cite{Boyer:1988hm}.  
$B$ must be conformally equivalent to one of the following:  a $T^4$
with a flat metric, a $K3$ with its hyper-K\"ahler metric, or  a
Hopf surface.  A Hopf surface is a quotient 
$\C^2 \backslash \{0\} / Z$, where $Z$ is a cyclic group of automorphisms
generated by
\begin{align}
\mathfrak{g} : (z_1,z_2) \mapsto (s z_1 +\lambda z_2^m, t z_2),
\qquad m \in \Z_{>0}, \quad s,t,\lambda \in \C,
\end{align}
with $0 < |s| \le |t| <1$ and $(t^m-s)\lambda = 0$\cite{Kodaira:1981cx}.
Each of these is a compact hyper-Hermitian surface diffeomorphic to $S^1\times S^3$.  It can be shown that a Hopf surface is locally
conformally hyper-K\"ahler~\cite{Boyer:1988hm}.  That is, in each coordinate patch there
exists a conformal rescaling of the metric that leads to a hyper-K\"ahler structure; however, these local rescalings cannot be patched
to a global function on the surface.

In fact, we can constrain $B$ further.  
As explained
in~\cite{Fu:2006vj,Becker:2006et}, $B = T^4$ requires 
the fibration and torsion to be trivial:  $X = T^6$,
and $H=0$. Thus, the ``simplest'' possibility for $B$ leads to
$N=4$ space-time SUSY. Can $B$ be a Hopf surface?
While all the requirements are locally satisfied, there is one
subtlety:  $B$ does not have a holomoprhically trivial canonical
bundle, so the background does not admit a single-valued dilaton.
The most direct way to see this is to consider the requirement of
conformal balance.  We have
\begin{align}
d (\omega \wedge \omega) = \betah \wedge \omega \wedge \omega \overset{?}{=}
2 d\vphi \wedge \omega\wedge \omega.
\end{align}
While $\betah$ is closed for all $B$, it is not exact for a Hopf surface.

For $B = K3$  all the conditions can be satisfied, and as expected we can construct two well-defined spectral flow operators via 
\begin{align}
\Sigma^2_\pm = \Omega^\pm_{\lambda\mu\nu}
\psi^\lambda\psi^\mu\psi^\nu, \qquad
\Omega^\pm = e^{2\vphi} \Omega_{K3} \wedge (\Theta^1 \pm i \Theta^2),
\end{align}
where we have for simplicity set $\tau = i$, and $\Omega_{K3}$ is
the holomorphic 2-form of the K3.  Moreover, as the analysis
of~\cite{Fu:2006vj,Becker:2009zx} shows, for $\omegah = e^{2\vphi} \omega_{K3}$, there exist solutions to the remaining
conditions of the Bianchi identity in~(\ref{eq:Bianchiomega}) and
HYM equations, provided the bundle is stable and satisfies the requisite topological conditions.  

Thus, within the assumptions of the NLSM approach, we can 
conclude that the only geometric heterotic compactification preserving exactly
$N=2$ space-time SUSY in four dimensions is a $T^2$ bundle over
$K3$ with primitive first Chern classes $F^I$, and torsion as determined
above. Furthermore, the base $K3$ must be equipped with a conformally 
hyper-K\"ahler metric, with the  dilaton proportional to the conformal factor.

\subsection{The gauge bundle}
Finally, we must discuss the SUSY constraints on the gauge bundle.
We take the connection to be
\begin{align}
\bcA = \bcAh_i dy^i + \bca_I \Theta^I,
\end{align}
so that $\bcAh(y)$ and $\bca_I(y)$ are invariant under the Kaluza-Klein gauge transformations of $\theta^I$.  This is a well-defined connection if
$\bcA$ transforms as a connection for a bundle $\Eh \to B$, while $\bca_I$ are
sections of $\ad(\Eh)$.\footnote{We will consider $\Eh$ that can be embedded
in the usual free fermion construction; more general bundles may require a more 
general world-sheet treatment~\cite{Distler:2007av}.}
The curvature has components
\begin{align}
\bcF_{ij} & = \bcFh_{ij} + F^M_{ij} \bca_M + A^M_j \bcDh_i \bca_M - A^M_i \bcDh_j \bca_M,
\nonumber\\
\bcF_{Ij} & = -\bcDh_j \bca_I - A^M_j \CO{\bca_M}{\bca_I}, \nonumber\\
\bcF_{IJ} &= \CO{\bca_I}{\bca_J},
\end{align}
where $\bcDh$ is the covariant derivative with respect to the ``base'' connection
$\bcAh$, and $\bcFh$ is its curvature.

The Fermi action will possess
$\cA_2\oplus\cA_4$ invariance if the curvature $\bcF$
satisfies the analog of (\ref{eq:Rsymcond2}):
\begin{align}
\cI^\nu_{\mu} \bcF_{\nu\lambda} +\bcF_{\mu\nu}\cI^\nu_\lambda = 0, \qquad
\cK^\nu_{a\mu} \bcF_{\nu\lambda} +\bcF_{\mu\nu}\cK^\nu_{a\lambda} = 0.
\end{align}
Expanding these conditions in the by now familiar base-fiber decomposition, we find
\begin{align}
\label{eq:Fcond}
\bcDh_j \bca_I = 0, \qquad 
\cKh^k_{a i} \bcFh_{k j} +\bcFh_{ik}\cKh^k_{aj} = 0.
\end{align}
The second condition in (\ref{eq:Fcond}) implies $\bcFh$ is primitive: $\omegah_a \wedge \bcFh = 0$, i.e.
$\bcAh$ must be a zero-slope HYM connection over $B$.  Since $N=1$ space-time SUSY also requires $\bcA$ to be a zero-slope HYM connection over $X$,
we have, using the primitivity of $\bcFh$ and $F^I$ on the base,
\begin{align}
0 = \omega\wedge\omega\wedge \bcF = \CO{\bca_1}{\bca_2} \omegah\wedge\omegah\wedge\Theta^1\wedge\Theta^2
\quad \implies \CO{\bca_1}{\bca_2} = 0.
\end{align}
Thus, the $\bca_I$ must be covariantly constant commuting elements of $\ad(\Eh)$.  A simple
way to satisfy the requirement is to pick the $\bca_I$ to be commuting constant matrices valued 
in the commutant of $G$ in $\GE_8\times \GE_8$ or $\SO(32)$.  In the familiar $K3\times T^2$ compactification, we recognize these as the commuting Wilson lines on the torus.

Having determined the constraints on the gauge connection, we can study the form of the Bianchi
identity in a little more detail.  Using either the holomorphic or the $H$-twisted connection to compute
$\tr R\wedge R$, a short computation shows that the Bianchi identity reduces to an equation on the
base
\begin{align}
d(\Hh + \frac{\alpha'}{4} \tr( \cdots) ) = \frac{\alpha'}{4} ( \tr \bcF^2 -\tr \Rh^2 
-\frac{4}{\alpha'} \cG_{IJ} F^I \wedge F^J),
\end{align}
where $\Rh$ is the Ricci curvature computed with the base metric $\gh$.  Integrating this
over the base we obtain an integrality condition on the torus metric and Wilson
lines (we take $\bca_I$ to be constant commuting matrices satisfying $\bca_I \bcFh = 0$):
\begin{align}
-\left(\frac{1}{\alpha'} \cG_{IJ} -\frac{1}{4}\tr(\bca_I \bca_J) \right) c_1^I \cdot c_1^J 
=24 - c_2(\Eh) + \frac{1}{2} c_1(\Eh)^2, 
\end{align}
where $c_1^I \cdot c_1^J = \int \frac{ F^I}{2\pi}\wedge \frac{ F^J}{2\pi}$.

\section{Discussion} \label{s:discussion}
In this note we have explored the consequences of $N=2$ space-time SUSY in the context of heterotic NLSMs.  Our basic result is that the $T^2$ principal fibrations over a K3 base already studied at length in the literature constitute the full class of solutions to the requirements of SUSY.  This is of course in marked contrast to the $N=1$ case,
where the class of geometries corresponds to complex $3$-folds with trivial canonical bundle.  Such geometries are surely abundant and fairly poorly understood;  the reader may consult~\cite{Fine:2009gh} for some recent constructions.

 As has been emphasized, for instance in~\cite{Becker:2006et,Becker:2009df}, an
 interesting and hopefully tractable class of $N=1$ backgrounds can be obtained by simply relaxing some of the requirements of $N=2$ SUSY.  Perhaps the most mild is to let $F^1 + \tau F^2$ have a (2,0) component over the base manifold.  A more drastic modification is to let the complex structure of the $T^2$ fiber vary holomoprhically over $B$.  In this situation, the local geometry and its relation to IIB/F-theory was studied in~\cite{Becker:2009df}.  Of course more dramatic modifications are
possible.  For instance, it is argued in~\cite{McOrist:2010jw}, there exist many non-geometric solutions, where the (complexified) volume of the $T^2$ is fibered over a base manifold.

We should emphasize again that the study of these backgrounds is complicated by the lack of large radius limit.  Since $\alpha'$ corrections are large, it is difficult to go beyond a qualitative description of the corresponding string vacua.  In the case of $N=2$ SUSY, some additional insight is obtained via the torsional linear sigma models of~\cite{Adams:2006kb,Adams:2009tt,Adams:2009zg}; some $N=1$ and $N=0$ theories can be constructed from these by taking an additional orbifold.\footnote{We thank A.~Adams for pointing out the additional orbifold possibility to us.}  In the $N=2$ case these linear models only make a (0,2) world-sheet SUSY manifest.  It would be interesting to develop descriptions that make manifest all six supercharges.  

A prime motivation for our investigation was to describe the (0,2)+(0,4) world-sheet supersymmetry explicitly.  While we were successful in finding this structure
even in the case of non-trivially fibered $T^2$, it remains a challenge to use this to constrain quantum corrections.  The difficulty is, of course, that our symmetries are not linearly realized on some familiar superspace.  It would be interesting to see to what extent the (0,2)+(0,4) supersymmetry can be given a superspace formulation.

While four-dimensional compactifications of heterotic strings provided the main motivation for this work, much of what was done in the previous section did not depend on the dimensions of the respective target spaces for (0,4) and (0,2) models.   It is well known that higher-dimensional (0,4) NLSMs
appear in the context of Calabi-Yau black holes. M5-branes wrapping very ample divisors give rise to such models, and their study has been important for the microscopic derivations of black hole entropy for half-BPS sectors in theories with eight supercharges.  More generic
(0,2) theories correspond to the largely unexplored quarter-BPS sector in theories with
eight supercharges.  The (0,2)+(0,4) structure could be a useful intermediate class of theories, so a natural question is what is the general class of models admitting this split.
There are clearly some simple generalizations of our construction,
but it may well be that this is just a small subset of the possible models.

An obvious generalization is to replace the $T^2$ fiber with a higher dimensional torus $T^{2k}$.  Of course in order for the full $T^{2k}$ to be non-trivially fibered, the base $B$ must have $H^2(B)$ is large enough to support the non-trivial fibration.  For 
instance on $B=K3$ we can have $k \le 9$.   One could also replace the base with
a hyper-Hermitian manifold of higher dimension.

A more interesting possibility would be to replace $T^2$ with a general even-dimensional compact Lie 
group $G$.\footnote{This should not be confused with constructions of~\cite{Distler:2007av}, where a fibered WZW
model is used to construct a left-moving current algebra.}
It is a classical result that each such $G$ admits a complex structure~\cite{Wang:1954cm}, and one might try to fiber this over a base $B$, thereby producing a fibered WZW model over $B$.  Unfortunately, this idea runs into
a simple problem for non-abelian $G$.  In order to implement our construction, we would
need to choose a complex structure on $G$ with a $G$-invariant Hermitian form.  It
is not hard to convince oneself that such a Hermitian form does not exist.   The difficulty is easily illustrated with $G = \SU(2)\times \SU(2)$.  Taking $e^a, f^a$
to be right-invariant $1$-forms on the two factors, it is easy to pick a complex structure, for example by choosing the (1,0)-forms to be
\begin{align}
\sigma^1 = e^1+i e^2, \quad 
\sigma^2 = e^3 + i f^1, \quad
\sigma^3 = f^2 + i f^3.
\end{align}
The corresponding Hermitian form is then
\begin{align}
\omega = e^1 \wedge e^2 + e^3 \wedge f^1 + f^2 \wedge f^3.
\end{align}
While perfectly well-defined, $\omega$ does not transform covariantly 
under the left $G$-action.  This should be compared to the recent work~\cite{Adams:2009av}, where a WZW model is non-trivially coupled to a gauged linear sigma model.  While it
is clear that in this fashion one can produce many new (0,2) theories, it might also be
interesting to study whether it is possible to realize new NLSMs with (0,2)+(0,4) supersymmetry as a  special case, for example
by working with GLSMs with manifest (0,4) supersymmetry, as in
~\cite{Witten:1994tz}.

\providecommand{\href}[2]{#2}\begingroup\raggedright\endgroup

\end{document}